\documentclass[12pt]{article}
\usepackage{epsfig}
\usepackage{amsmath}
\usepackage{amssymb}

\let \to=\rightarrow
\newcommand{\cp}{{\mathbb{C}{\mathbf{P}^{2}}}}
\newcommand{\be}{\begin{equation}}
\newcommand{\ee}{\end{equation}}

\begin{document}

\baselineskip=16pt

\setcounter{footnote}{0}
\setcounter{figure}{0}
\setcounter{table}{0}

\begin{titlepage}

\begin{center}
\vspace{1cm}

{\Large \bf  Nonperturbative Instability of  $\text{AdS}_{5} \times {\bf S}^{5}/\mathbb{Z}_{k}$}

\vspace{0.8cm}

{\bf Gary T.  Horowitz$^1$, Jacopo Orgera$^1$, Joe Polchinski$^2$}

\vspace{.5cm}

{\it $^1$ Department of Physics, UCSB, \\ Santa Barbara, California 93106, USA}

{\it $^2$ Kavli Institute for Theoretical Physics \\Santa Barbara, California 93106, USA}

\end{center}
\vspace{1cm}

\begin{abstract}
\medskip
We study the AdS/CFT correspondence with boundary conditions ${\rm AdS}_{5} \times {\bf S}^{5}/\mathbb{Z}_{k}$, where the $\mathbb{Z}_{k}$ acts freely  but breaks all supersymmetry. While there are closed string tachyons  at small 't Hooft coupling, there are no tachyons at large coupling.  Nevertheless, we show that there is a nonperturbative instability directly analogous to the decay of the Kaluza-Klein vacuum. We discuss the implications of this instability for the strongly coupled dual field theory, and compare with earlier studies of this theory at weak coupling.

\end{abstract}

\bigskip
\bigskip

\end{titlepage}
\baselineskip=18pt
\setcounter{equation}{0}

\section{Introduction}

By orbifolding on the $SU(4)$ R-symmetry of ${\mathcal N} = 4$ SYM and its $\text{AdS}_5\times {\bf S}^5$ dual one obtains a rich set of AdS/CFT duals with reduced supersymmetry~\cite{KSorb,Lawrence:1998ja,Oz:1998hr}.  However, when the supersymmetry is broken completely there are a number of potential instabilities and sources of conformal symmetry breaking, and the full picture is not yet clear.

At small 't Hooft coupling $\lambda$, there is conformal symmetry breaking in a twisted sector even at the planar level~\cite{Adams:2001jb,Dymarsky:2005uh,Dymarsky:2005nc}.  In all examples studied thus far, some of the couplings have Landau poles both in the UV and in the IR.  The IR pole has been shown, at least in many cases, to give rise to spontaneous breaking of a twist symmetry via the Coleman-Weinberg mechanism.  The UV pole presumably has the same implication as it does in $\phi^4$ theory, that the field theory can only be regarded as effective up to some maximum energy scale.  There are also some interesting parallels in the noncommutative case~\cite{Armoni:2003va}.

At large $\lambda$, if the orbifold has fixed points (i.e.\ is non-freely acting), then there are twisted sector closed string tachyons~\cite{Adams:2001jb,Dymarsky:2005uh,Dymarsky:2005nc,Armoni:2003va}.  These have no stable ground state, but instead quickly consume the entire space~\cite{APS}.  (For a related discussion, see \cite{Morrison:2004fr}.)  When the orbifold is freely acting, the twisted sector closed strings are stretched over lengths of order the AdS radius 
\be
R = \lambda^{1/4}\alpha'^{1/2}
\ee
 and so there are no tachyons at sufficiently large $\lambda$.  

Nevertheless one still expects an instability~\cite{FabHor,Adams:2001jb,APS}.  When a closed string tachyon is stretched far enough its mass term becomes positive and the tree level instability is absent, but decay may still take place via tunneling.  This tunneling has a natural interpretation~\cite{FabHor,Horowitz:2000gn} as a Kaluza-Klein (KK) `bubble of nothing' instability~\cite{Witten:1981gj}, with the direction on which the closed string is wrapped playing the role of the KK direction.  In this paper we work out the details and implications of this decay.

There have been several previous discussions of `bubble of nothing' instabilities in the context of AdS/CFT \cite{Birmingham:2002st,Balasubramanian:2005bg}. However, those examples require a modification of the conformal metric on the boundary of AdS. In particular, they describe instabilities of gauge theory on ${\bf S}^1\times \text{dS}_3$. We will keep the standard asymptotic AdS geometry and just consider an orbifold acting on ${\bf S}^5$.

We will focus on a simple example, in which the generator of the orbifold group $\mathbb{Z}_{k}$ acts by an equal rotation by $2\pi/k$ in each of the three transverse planes.  For reasons to be described in the next section, $k$ must be odd and the generator must include an additional $-1$ acting on spacetime fermions.  We can think of ${\bf S}^5$ as a Hopf fibration of ${\bf S}^1$ over $\mathbb{C}{\mathbf{P}^{2}}$.  In the parent theory, both the circle and the base have a radius of order $R$.  The orbifold identifies the ${\bf S}^1$ factor modulo $\mathbb{Z}_{k}$, reducing its radius to $R/k = (4\pi g_{\rm s} N)^{1/4} \alpha'^{1/2}/k$.  To avoid perturbative tachyons,  we thus require $g_{\rm s} N \gg k^4$.

If we consider the limit $N \to \infty$ with $N/k^4$ and $g_{\rm s}$ held fixed,  the space becomes flat with the radius of the KK circle fixed, and we simply go over to the ordinary Kaluza-Klein instability~\cite{Witten:1981gj}.  Thus we can conclude that this instability will be present for sufficiently large $k$.  However, the usual KK vacuum has a one loop Casimir energy which can cause the circle to shrink in size. In particular, it can shrink to the string scale, giving rise to closed string tachyons, long before the nucleation of a `bubble of nothing'.  In our case, the five-form flux contributes an energy density of order $N^2/R^{10}$ while the Casimir energy associated with a small circle of size $R/k$ is of order $(k/R)^{10}$. Thus, provided 
\be\label{casimir}
N^2 \gg k^{10}
\ee
 the Casimir energy will be negligible and not cause the circle to shrink. The dominant decay mode will be nucleation of a bubble of nothing.  Note that (\ref{casimir}) is violated in the flat limit above. 
  
  In order to determine the lower limit on $k$ we will need to study the solutions numerically.  We find that the instability persists down to the lowest relevant value, which is $k=5$; for $k=3$ the orbifold is actually supersymmetric.  The supersymmetric case $k =3$ is interesting.  In the usual case~\cite{Witten:1981gj} the KK instability is topologically forbidden with supersymmetric boundary conditions since the only spin structure on the bounce spacetime  breaks supersymmetry.  In our case, there is a spin structure compatible with supersymmetry, but  the bounce is forbidden dynamically.

We consider both bounces that are fully localized in  nine dimensions, and bounces that are smeared over the $\mathbb{C}{\mathbf{P}^{2}}$.  The localized bounces have lower action, but the smeared bounces are more symmetric and more easily studied numerically.  For the smeared bounces we find that there must be an RR source, so that the bounce is essentially a combination of a KK bounce and a D-instanton.

In Sec.~2 we describe the orbifold and the smeared bounce.  We show that a RR source is needed, and obtain the equations of motion.  In Sec.~3 we show that an approximate solution can be obtained at large $k$, where the KK circle is small, in terms of analytic solutions in three overlapping regimes.  We also describe the localized bounce in this limit, and discuss a puzzle with the spin structure.  In Sec.~4 we discuss the numerical integration of the smeared bounce solution, discussing in particular how to fix parameters to match the short- and long-distance asymptotics.  We find that a solution exists for $k > 3$.
In Sec.~5 we discuss the physical implications of our result.  We argue that the strongly coupled theory can only be defined in an effect sense, with both IR and UV cutoffs.  The symmetry breakings in the weak and strong coupling limits are similar. 

\setcounter{equation}{0}

\section{The instability}\label{kkinst}

\subsection{The orbifold}

Consider first the orbifold of $\text{AdS}_5\times {\bf S}^5$ by the rotation
\begin{equation}
g = e^{2\pi i (J_{45} + J_{67} + J_{89})/k}\ .
\end{equation}
This has no fixed points on ${\bf S}^5$, but if we raise it to the $k^{\rm th}$ power we get
\begin{equation}
g^k = e^{2\pi i (J_{45} + J_{67} + J_{89})} = (-1)^{\bf F}\ .
\end{equation}
This acts trivially on the spatial coordinates, but because we have rotated by $2\pi$ in an odd number of planes, it multiplies every spacetime fermion by $-1$. Thus the projection by $g^k$ removes all fermions from the theory and leaves the type 0 theory, with a bulk tachyon.  Thus we consider instead
\begin{equation}
g' = (-1)^{\bf F} e^{2\pi i (J_{45} + J_{67} + J_{89})/k}\ ,
\label{gprime}
\end{equation}
for which
 \begin{equation}
g'^k = (-1)^{(k+1) \bf F}\ .
\end{equation}
For $k$ even this again produces the type 0 theory, but for $k$ odd it is the identity and we get the desired $\mathbb{Z}_{k}$ orbifold of the IIB theory without fixed points.
The rotation $g'$ acts on the spinors in the ${\bf 4}$ as diag$(-e^{\pi i/k}, -e^{\pi i/k}, -e^{\pi i/k}, -e^{-3\pi i/k} )$.  For $k > 3$, all of these phases are nontrivial and the supersymmetry is completely broken.  For $k=3$ an ${\cal N}=4$ supersymmetry survives.

On the gauge theory side this orbifold has been studied in Ref.~\cite{Dymarsky:2005uh}.  Our interest is the string theory description, at large $\lambda$.  In the sector twisted by $g'$, the string has a negative zero point energy $2(-1 + 3/k)/\alpha'$ (see e.g.~Ref.~\cite{Dymarsky:2005nc}) but stretches over a minimum distance $2\pi R/k$, giving a ground state with
\begin{equation}
\alpha' M^2 = \frac{R^2}{\alpha' k^2} + \frac{2(3-k)}{k} 
= \frac{\lambda^{1/2}}{k^2} + \frac{2(3-k)}{k}\ .
\end{equation}
This is always positive at large enough $\lambda$, so there is no tree level instability.

\subsection{The smeared bounce solution}

We will show that $\text{AdS}_{5} \times {\bf S}^{5}/\mathbb{Z}_{k}$ has a nonperturbative instability analogous to the decay of the Kaluza-Klein vacuum, $\mathbb{R}^n \times {\bf S}^1$ \cite{Witten:1981gj}. Recall that Witten showed that (with antiperiodic fermions around the circle) there was a nonzero probability to nucleate a `bubble of nothing'.  In other words, the Kaluza-Klein circle pinches off at a finite radius in $\mathbb{R}^n$ producing a minimal sphere called the bubble. There is no spacetime inside this bubble, which rapidly expands out and hits null infinity. The instanton describing the nucleation of this bubble can be simply obtained by analytic continuation of the $n+1$ dimensional Schwarzschild solution. 

The Kaluza-Klein direction here is generated by the $J_{45} + J_{67} + J_{89}$ that appears in the  $\mathbb{Z}_{k}$ generator.  The ${\bf S}^{5}/\mathbb{Z}_{k}$ is a Hopf fibration of this ${\bf S}^1$ over $\mathbb{C}{\mathbf{P}^{2}}$.   Thus we can think of $\text{AdS}_{5} \times {\bf S}^{5}/\mathbb{Z}_{k}$ as a KK compactification down to $\text{AdS}_{5} \times \mathbb{C}{\mathbf{P}^{2}}$, with a KK gauge field on the $\mathbb{C}{\mathbf{P}^{2}}$ coming from the fibration.  The metric of Euclidean $\text{AdS}_{5} \times {\bf S}^{5}$  is
\begin{equation}
ds^2 =   \frac{ dr^2}{1+{r^2\over R^2}} + r^2 d\Omega_4^2 + R^2 \left [ds^2_{\mathbb{C}{\mathbf{P}^{2}}} + (d\chi + A)^2 \right ] \label{ads5s5}
\end{equation}
The first two terms are the metric of $\text{AdS}_{5} $ in a convenient coordinate system, while $\chi$ is the coordinate on the fiber and $A$ a gauge connection on $\mathbb{C}{\mathbf{P}^{2}}$.
Using the coordinates
\begin{eqnarray}
z^1 &=& e^{i(\phi_1 + \chi)}\cos\theta \ ,\nonumber\\
z^2 &=& e^{i(\phi_2 + \chi)}\sin\theta \cos\psi \ ,\nonumber\\
z^3 &=& e^{i\chi}\sin\theta \sin\psi  \ ,
\end{eqnarray} 
which satisfy $z^i \bar z^i = 1$ and $d\Omega_5^2 = dz^i d\bar z^i$, we have
\begin{eqnarray}
ds^2_{\mathbb{C}{\mathbf{P}^{2}}}  &=& \sum_{a=1}^4 e_a e_a\ ,\nonumber\\
d\chi + A &=& e_5\ .
\end{eqnarray} 
Here
\begin{eqnarray}
e_1 &=& d\theta \ , \nonumber\\
e_2 &=& \sin\theta\, d\psi \ ,\nonumber\\
e_3 &=& \cos\theta\sin\theta (d\phi_1 - \cos^2\chi \,d\phi_2) \ ,\nonumber\\
e_4 &=& \sin\theta\cos\psi\sin\psi \, d\phi_2 \ , \nonumber\\
e_5 &=& d\chi+ \cos^2 \theta\, d\phi_1 + \sin^2 \theta\cos^2\psi \, d\phi_2\ .
\label{5bein}
\end{eqnarray} 
On the original ${\bf S}^{5}$ the periodicity of $\chi$ is $2\pi$, so the $\mathbb{Z}_{k}$  orbifold is constructed simply by reducing this to $2\pi/k$.  Note that this commutes with the $U(3)$ symmetry that acts on the $z^i$.

We consider first bounces that are smeared on the $\mathbb{C}{\mathbf{P}^{2}}$.  The bounce on flat $\mathbb{R}^5$ has $SO(5)$ symmetry, with the radius of the KK circle pinching to zero at finite $r$.  Thus we replace the metric~(\ref{ads5s5}) with the most general metric preserving this $SO(5)$ symmetry and the $U(3)$ symmetry of the $z^i$,
\begin{equation}
ds^2 = \rho(r) dr^2 + f(r) d\Omega_4^2 + g(r) ds^2_{\mathbb{C}{\mathbf{P}^{2}}} +  h(r)(d\chi + A)^2\ .
\label{metric}
\end{equation}
We could fix one of the functions by a coordinate redefinition of $r$, e.g.~$\rho(r) = 1$ or $f(r) = r^2$, but it is convenient to be more general.  For the KK decay we are looking for a solution that matches onto $\text{AdS}_{5} \times {\bf S}^{5}/\mathbb{Z}_{k}$ at large $r$ but where $h(r)$ goes to zero at a radius where $f(r)$ is still nonzero.

For this smeared instanton, the Lorentzian evolution after the bubble nucleation is easily obtained by analytically continuing ${\bf S}^4$ in Euclidean $\text{AdS}_{5}$ to ${\rm dS}_4$. In other words, the $SO(5)$ symmetry of the instanton translates into a $SO(4,1)$ symmetry of the Lorentzian solution. Asymptotically, the solution looks like $\text{AdS}_{5}$ in de Sitter slices. These slices stay outside the light cone of the origin of $\text{AdS}_{5}$. Thus the bubble of nothing accelerates out and reaches infinity in finite (global) time. This is clearly a serious instability.

For the usual KK bubble, the geometry is smooth at the radius $r_0$ where the geometry pinches off.  This would require that $h(r)$ vanish quadratically; more precisely, 
\begin{equation}
h(r) \approx \rho(r_0) k^2 (r - r_0)^2
\end{equation}
for $\chi \cong \chi + 2\pi/k$.  In the present case, however, there is a complication.  
There is a nonzero flux on the ${\bf S}^{5}/\mathbb{Z}_{k}$
\be\label{fluxquantum}
\frac{1}{(2\pi)^3 \alpha'^2} \int_{S^5/Z_k} F_5 = 2\pi N/k\ .
\ee
However, in the bounce geometry, ${\bf S}^{5}/\mathbb{Z}_{k}$  is the boundary of a six-manifold, because the KK circle is the bou   ndary of a disk, and so we should have $\int_{S^5/Z_k} F_5 = \int_{M_6} dF =0$.  Thus there must be a $N/k$ D3-brane instantons wrapped on the ${\bf S}^4$ at $r = r_0$, and smeared on the  $\mathbb{C}{\mathbf{P}^{2}}$.  Therefore the bounce geometry will not be smooth, but will have the singularity of a smeared $D3$.

We also can think of this as follows.  We can imagine charging up the $F_5$ flux by dropping spherical D-branes in from the asymptotic ${\bf S}^3$ of the $\text{AdS}_{5}$.  Without the bubble, these would just drop to the center and annihilate, but when there is a bubble of nothing in the center they can only drop down and wrap bubble.  In the continued Lorentzian geometry, these then expand with the bubble.\footnote{Without the bubble, such spherical D3-branes are energetically forbidden~\cite{Seiberg:1999xz}: this energy is dual to the positive curvature term on the Coulomb branch of the gauge theory on ${\bf S}^3$.}

The $SO(5)$ symmetry and the  imaginary selfduality 
condition $F_5 = i \ast F_5 $ determine the form
\begin{equation}
 F_5 = \xi(r)(\varepsilon_5 \oplus i \ast \varepsilon_5)\ ,
  \end{equation}
where $\varepsilon_5$ is the volume form on the $ {\bf S}^{5}/\mathbb{Z}_{k}$ part of the geometry. The equation of motion and the Bianchi identity for $F_5$ then determine
\begin{equation}
\xi(r) = \frac{4 R^4}{g_s g(r)^2 h(r)^{1/2}}\ ,
\end{equation}
with the normalization fixed by (\ref{fluxquantum}).
The Einstein equations are then given by
\begin{equation}
R_{MN} = \pm \frac{4 R^8}{g^4 h}g_{MN}\ ,
\end{equation}
where the sign is $-$ for the hyperbolic part of the geometry, and $+$ for spherical part. There are four nontrivial equations
\begin{equation}
\begin{split}\label{EOM}
E_{\rho} & = \frac{4 f' g'}{f g} + \frac{f' h'}{f h} + \frac{g' h'}{g h} + \frac{3 f'^2}{2f^2} + \frac{3 g'^2}{2 g^2} + \frac{2 h \rho}{g^2} - \frac{6 \rho}{f} - \frac{12 \rho}{g} + \frac{4 \rho  R^8}{g^4 h} = 0\ ,\\
E_{f} & =  -\frac{f''}{2 \rho} - \frac{f'^2}{2 f \rho} - \frac{f' g'}{g \rho} - \frac{f' h'}{4 h \rho} + \frac{f' \rho '}{4 \rho^2}+ 3 + \frac{4 f R^8}{g^4 h} = 0\ ,\\
E_{g} & =  - \frac{g''}{2 \rho} - \frac{g'^2}{2 g \rho} - \frac{f' g'}{f \rho} - \frac{g' h'}{4 h \rho} + \frac{g' \rho '}{4 \rho ^2} - \frac{2 h}{g} + 6 - \frac{4 R^8}{g^3 h} = 0\ ,\\
E_{h} & = - \frac{h''}{2 \rho} + \frac{h'^2}{4 h \rho} - \frac{f' h'}{f \rho} - \frac{g' h'}{g \rho} + \frac{h' \rho '}{4 \rho^2} + \frac{4h^2}{g^2} - \frac{4 R^8}{g^4} = 0\ ,
\end{split}
\end{equation}
where we have defined
\begin{equation}\label{normEOM}
\begin{split}
R_{rr} + \frac{4 R^8}{g^4 h}g_{rr} - \frac{\rho}{2}\mathcal{R} &=
E_{\rho} \ ,\\
R_{MN} + \frac{4 R^8}{g^4 h}g_{MN}
&=  E_{f}g_{MN}/f\ ,\quad M,N{\ \|\  } {\bf S}^4\ ,  \\
R_{MN} - \frac{4 R^8}{g^4 h}g_{MN}
&= E_{g}g_{MN}/g \ ,  \quad M,N\ \|\ \mathbb{C}{\mathbf{P}^{2}}\ , \\
R_{\chi\chi} - \frac{4 R^8}{g^4 h}g_{\chi\chi} &= E_{h} \ .
\end{split}
\end{equation}
We have subtracted $\rho\mathcal{R}/2$ from the first component of Einstein's equation (where $\mathcal{R}$ is the scalar curvature) to make it first-order.  As usual only three of these are independent due to the Bianchi identity,
and so they determine three of the four functions $\rho(r)$, $f(r)$, $g(r)$, and $h(r)$, with the fourth being a gauge choice.

\setcounter{equation}{0}

\section{Analytic approximations}

\subsection{Large $k$ analysis} 

The Einstein equations~(\ref{EOM}) apparently cannot be solved analytically, so we will first consider a regime, large $k$, where we can solve them approximately, and then go to smaller values of $k$ numerically.  At large $k$, the radius of the KK circle is $R/k$, parametrically smaller than the radii of the other factors.  The curvature of the KK bubble is correspondingly of order $k^2$ times the curvature of the $\text{AdS}_{5} \times {\bf S}^{5}/\mathbb{Z}_{k}$.  Thus we can ignore the latter curvature near the bubble, and graft the flat-spacetime bubble solution into the geometry.  At distances large compared to the size of the bubble, on the other hand, we can approximate the solution by the unperturbed $\text{AdS}_{5} \times {\bf S}^{5}/\mathbb{Z}_{k}$.  These approximations overlap at distances large compared to $R/k$ and small compared to $R$.

A third approximate form is needed because of the smeared D3 RR source noted above.   The bulk AdS curvature produced by the RR field is smaller than that of the bounce, but as one approaches the source there will be a singularity in this field.  Thus there is a third regime, near the D3-branes.  
This regime begins close to the D3-branes, much closer than the scale $R/k$, and so we can treat the source as being in a locally flat spacetime. 

In the asymptotic regime we thus have
\begin{equation}\label{BG}
\rho_{\text{I}} = 1\,, \quad f_{\text{I}} = R^2\sinh^2 (r/R)\,, \quad g_{\text{I}} = h_{\text{I}} = R^2\ ,
\end{equation}
which is the Euclidean $\text{AdS}_{5} \times {\bf S}^{5}/\mathbb{Z}_{k}$ geometry with a convenient choice of the coordinate $r$ (different from Eq.~(\ref{ads5s5})).
In the bubble regime, the approximate solution is
\begin{equation}\label{BH}
\rho_{\text{II}} = \frac{1}{H(r)}\,, \quad f_{\text{II}} = r^2\,, \quad g_{\text{II}} = R^2\,, \quad  h_{\text{II}} = R^2 H(r)
\end{equation}
where 
\begin{equation}
H(r) = 1-\frac{r_0^3}{r^3}\ .
\end{equation}
The geometry in the AdS plus $\chi$ directions is essentially the flat spacetime KK bounce (for five large spacetime dimensions), which is just the six-dimensional Euclidean black hole~\cite{Witten:1981gj}. It differs only by the off-diagonal $A$ term in the metric (\ref{metric}). However this plays no role on scales of the bounce: for large $k$, $A$ is approximately constant on scales of order $R/k$ and can locally be absorbed by shifting $\chi$.

The coordinate $r$ starts from $r_0$, and regularity at $r=r_0$ determines $r_0$ as follows:
A change of variables, $r - r_0 = \frac{3 \tilde{r}^2}{4 r_0}$,  puts us in a gauge where the radial coordinate starts at the origin and, for small $\tilde{r}/r_0$, is the proper distance. In the $\tilde r$ coordinate the metric coefficients are
\begin{equation}\begin{split}\label{nearoriginBH}
\tilde\rho_{\text{II}} &= 1  + \mathcal{O}({\tilde{r}}/{r_0})^2\ , \quad \tilde f_{\text{II}} = r_0^2 \left\{1  + \mathcal{O}({\tilde{r}}/{r_0})^2\right\}\ ,\\
\tilde g_{\text{II}} &= R^2\ ,\quad \tilde h_{\text{II}} =  \left(\frac{3 R}{2 r_0}\right)^2 \tilde{r} ^2  \left\{1  + \mathcal{O} ( {\tilde{r}}/{r_0})^2\right\}\ .
\end{split}\end{equation}
Since $\chi$ has periodicity $2\pi/k$, in order to avoid conical singularities we need to impose 
\begin{equation}\label{r0}
 r_0 = \frac{3 R}{2k} \ .
\end{equation}
The limit $r \ll  R$ of the metric \eqref{BG} agrees with the limit $r_0 \ll  r$ of \eqref{BH}, so we have overlap between the two approximate solutions.  

The final term in each of the Einstein equations~(\ref{EOM}) comes from the RR flux. For $\tilde r \sim r_0$, this term is smaller than the largest contributions to  each equation by a factor of $(r_0/R)^2 \sim 1/k^2$. So our vacuum solution is a good approximation. However,  as $\tilde r \rightarrow 0$ this term dominates all the others (since $\rho$ diverges and $h$ vanishes in this limit). So we must modify the solution near the origin as argued above.

The required source is a configuration of (Euclidean) D3-branes wrapping the ${\bf S}^4$ in $\text{AdS}_5$ and smeared over the $\cp$. Near the source, the solution should reduce to the usual flat brane result.  Since there are only two orthogonal directions (we are smearing over four directions), the relevant harmonic function is a log. Hence we expect the line element \eqref{nearoriginBH} to be warped near the origin as follows
\begin{equation}\begin{split}\label{harmonic}
& \rho_{\text{III}} = a \sqrt{- \ln (\tilde{r} /r_{\ast}) }
\ ,\quad\quad f_{\text{III}} = \frac{b r_0^2}{\sqrt{-\ln (\tilde{r} /r_{\ast})}}\ ,\\
& g_{\text{III}} = c R^2 \sqrt{- \ln (\tilde{r}/r_{\ast})}\ , \quad\quad
h_{\text{III}} =   a k^2 \tilde{r}^2  \sqrt{- \ln (\tilde{r} /r_{\ast}) }\ .
\end{split}\end{equation}
We have introduced  extra constants $a,b,c$ multiplying the metric in the localized, wrapped, and smeared directions respectively, as well as the integration constant $r_*$ in the log.  In a flat background this would be a solution, with three combinations of constants being varied by rescaling the three sets of coordinates, and the fourth determined by the equations of motion in terms of the density of D3 sources.  In the present case there is only one coordinate freedom $\tilde r$, and two of the constants are determined by matching onto the metric in region~II.

We fix the gauge freedom by setting $a=c$,  and then find that the leading terms in all of the equations of motion are cancelled by the choice
\begin{equation}
a^2 = c^2 = \frac{8 r_0}{3 R} = \frac{4}{k}\ .
\end{equation}
In order to complete the construction we need to determine the parameters $b$ and $r_*$ by matching the region II and region III solutions~(\ref{nearoriginBH}) and~(\ref{harmonic}).  Taking 
$\tilde{r} \ll r_0$ but $|\ln(r_0/\tilde r)| \ll k$, we find that the solutions match for
\begin{equation}
a^2 \ln(r_*/r_0) = 1\ ,\quad b = 1/a\ . \label{match}
\end{equation}
This implies that $r_* = r_0 e^{k/4}$ is exponentially large, so the logarithm is slowly varying.

In Section~3 we will verify this construction by numerical integration outward, interpolating a line element with the behavior \eqref{harmonic} near the singularity to one that has the asymptotic behavior of \eqref{BG}.   This will also allow us to extend the result to more general values of $k$.

The Einstein-Hilbert action for the smeared instanton is larger than the background by an amount of order the action for the Euclidean six dimensional black hole:
\be\label{ratesm}
B_{\rm smeared,EH} \sim {r_0^4\over G_6} \sim {R^8\over k^4 G_{10}} \sim {N^2\over k^4}\ .
\ee
We must also include the contribution of the D3-branes wrapped on the minimal sphere,
\be
B_{\rm smeared,D} \sim \frac{N}{k} \tau_{\rm D3} r_0^4 \sim  {N^2\over k^5}\ ,
\ee
but we see that this is parametrically smaller in the large-$k$ regime.

\subsection{The localized bounce}

In addition to the instantons constucted above, there are instantons which are localized on $\cp$.  These are less symmetric: the geometry depends both on the distance in $\text{AdS}_{5}$ and the distance along $\cp$.  We will therefore not write these metrics as explicitly, but when $k$ is large we can again describe their approximate form.   Since the size of the KK circle is small compared to the other factors,  
 the spacetime locally resembles $\mathbb{R}^9\times {\bf S}^1$ which has the usual Kaluza-Klein instability.    The instanton is obtained by  analytic continuation of the ten dimensional Schwarzschild solution. The parameter $r_0$ in this solution is fixed by requiring that the radius of the Euclidean time direction is $R/k$; this requires $r_0 \sim R/k$.  For $r\gg r_0$ this solution also approaches $\mathbb{R}^9\times {\bf S}^1$. Hence, for $R/k \ll r\ll R$, both the ten dimensional instanton and $\text{AdS}_{5} \times {\bf S}^{5}/\mathbb{Z}_k$ resemble  $\mathbb{R}^9\times {\bf S}^1$  and can be joined together.  The symmetries match as follows. The ten dimensional Schwarzschild instanton has eight dimensional spheres of spherical symmetry. Write the metric on ${\bf S}^8$ as
 \be
 d\Omega_8^2 = d\theta^2 + \sin^2\theta d\Omega_3^2 + \cos^2\theta d\Omega _4^2
 \ee
 The ${\bf S}^4$ symmetry remains and matches onto the spheres in Euclidean $\text{AdS}_{5}$. The ${\bf S}^3$ reflects the approximate rotational symmetry near a point in $\cp$.

The behavior of $F_5$ for the localized instanton can easily be obtained from \cite{Horowitz:2000kx} where the solution for $F_5$ near a small ten dimensional Schwarzschild black hole in $AdS_{5} \times {\bf S}^{5} $ was found.  If one analytically continues $t\to i\chi$ in that solution, $F_5$ becomes imaginary self-dual and remains nonsingular.

The localized instanton gives rise to a Lorentzian geometry which cannot be described explicitly, but is expected to be qualitatively similar to the smeared case. On scales smaller than the AdS radius, the bubble accelerates out as in the standard Kaluza-Klein instability. When it reaches the AdS scale, it is moving close to the speed of light. It is then plausible that it will not be affected much by the weak curvature at $r\approx R$ and will continue to expand in both the $\cp$ and $\text{AdS}_5$ directions at this speed wiping out the entire space. There are some examples of bubbles of nothing which stop accelerating after a while \cite{Aharony:2002cx}, however these typically occur when the Kaluza-Klein circle becomes large asymptotically. In our case, the size of the circle at infinity is finite  so we expect the acceleration to continue.

Note that, unlike the smeared bounce, the localized bounce has no D-brane source, as the ${\bf S}^5/\mathbb{Z}_k$ in this case is not a boundary.  In terms of the picture of charging up the global $\text{AdS}_5$, both the spherical D3-branes and the bounce are localized on the ${\bf S}^5$, so the D3-branes can just go past the bounce and collapse completely.  However, as the bubble expands, and as bubbles merge, they may consume the entire $\cp$ before reaching the boundary of $\text{AdS}_5$, so that holes in $\text{AdS}_5$ appear with  ${\bf S}^3 \times \mathbb{R}$ boundary. If this happens, spherical D3-branes must be produced since ${\bf S}^{5}/\mathbb{Z}_k$ becomes the boundary of a six manifold  and the flux requires explicit sources as before.

Since the  localized instanton has no source, its action is simply given by
\be\label{local}
B_{\rm local} \sim {r_0^8\over G_{10}} \sim {N^2\over k^8}
\ee
Note that for large $k$, this is much smaller than the action for the smeared instanton (\ref{ratesm}). However, it is still large in the limit that the one-loop Casimir forces can be neglected (\ref{casimir}).

\subsection{A spin structure puzzle}

There are two choices for the periodicity of fermions around a KK direction.  The usual KK bubble has a spin structure only if the fermions are antiperiodic~\cite{Witten:1981gj}.  Supersymmetric KK compactifications are therefore stable, because the fermions are periodic and so the bubble is topologically forbidden.

In our case, we are faced with the following puzzle: The bounce instanton has a unique spin structure since the circle at infinity bounds a disk. The question is whether this spin structure is compatible with the asymptotic behavior of fermions induced by the orbifold~(\ref{gprime}). In the limit of large $k$, 
 the fermions are clearly antiperiodic due to the factor of $(-1)^{\bf F}$. However,  we have seen that when we get down to $k=3$ the orbifold becomes supersymmetric. Previous experience would suggest that the bubble is topologically forbidden for $k=3$, yet allowed for large $k$.  This seems implausible, and indeed we now show that at $k=3$ there is a spin structure on the bounce that approaches the supersymmetric spin structure at long distance. 

Consider the zehnbein
\begin{eqnarray}
\tilde e_a &=& g^{1/2}(r)  e_a\ , \quad a = 1,2,3,4\ ,\nonumber\\
\tilde e_5 &=&  \cos k\chi\, dr - h^{1/2}(r) \sin k\chi\, e_5 \ ,\nonumber\\
\tilde e_6 &=&  \sin k\chi\, dr + h^{1/2}(r) \cos k\chi\, e_5 \ , \nonumber\\
\tilde e_a &=& f^{1/2}(r) \sigma_{a-6}\ ,\quad  a = 7,8,9,10
\label{space}
\end{eqnarray} 
for the metric~(\ref{metric}) with the $r$ coordinate such that $\rho=1$.
Here the $e_a$ are as in Eq.~(\ref{5bein}), and $\sigma_a$ is a vierbien for the unit $S^4$.
We will focus on a disk where the four $\cp$ coordinates are fixed.\footnote{The fact that $\cp$ has no spin structure is not an issue, because the KK gauge field from the twisting of the KK circle offsets this.}
If $h^{1/2}(r)$ vanishes as $k (r - r_0)$ for $r \to r_0$, while $f$, and $g$ remain nonzero, then this space is smooth at $r=r_0$.  The frame has been chosen so as to be smooth at the origin: in Cartesian coordinates $x = (r-r_0)\cos k\chi$, $y = (r-r_0) \sin k\chi$
it is just $dx,\, dy$ there.  Thus, it is globally defined on the disk, and fermions must be {\it periodic} under $\chi \to \chi + 2\pi/k$.  The nontrivial terms in the spin connection are
\begin{eqnarray}
\omega_{13} &\cong& \omega_{24} \cong d\chi\ , \nonumber\\
\omega_{56} &\cong& [h^{1/2}(r)' - k] \,d\chi\ , 
\end{eqnarray} 
where $\cong $ signifies that we keep only terms tangent to the disk.

Let us consider first a space which is asymptotically $\mathbb R^6$, so $h^{1/2}(r)$ approaches $r$ at large $r$.  The condition for supersymmetry is a constant spinor asymptotically on the disk:
\begin{equation}
\partial_\mu \psi = -i (\omega_{\mu 13} s_1 + \omega_{\mu 24} s_2 + \omega_{\mu 56} s_3) \psi\ ,
\label{conspin}
\end{equation}
with spin components $s_{1,2,3} = \pm \frac{1}{2}$ in the indicated planes.  Thus, at long distance,
\begin{equation}
\partial_\chi \psi = -i (s_1 + s_2 + [1 - k] s_3 ) \psi \ .
\end{equation}
The condition that $\psi$ be periodic on the orbifold is then
\begin{equation}
s_1 + s_2 + [1 - k] s_3 = 0 \bmod k\ . \label{flatsusy}
\end{equation}
For $k=1$ this holds for all choices of $s_i$.  The only other solution is $k=3$, when all the $s_i$ have the same sign.

The background of interest to us is a warped version of  the above, for which $h^{1/2}(r) = R$ at long distance.  However, we cannot apply the condition~(\ref{conspin}) directly, because the  five-form field strength also appears in the supersymmetry condition.  In fact, the effects of the warping and five-form just offset (for one linear combination of the IIB supersymmetries), so it is the flat spacetime condition~(\ref{flatsusy}) that is relevant.

It follows that the bounce spacetime is topologically allowed even in the case $k = 3$ where there is asymptotic supersymmetry.  Thus the stability of the supersymmetric orbifold must have a different explanation, and we will see in the next section that it is dynamical: there is no solution in this case.

\setcounter{equation}{0}

\section{Numerical integration}

To reach smaller values of $k$ we must resort to numerical integration.  We will study the smeared solution because of its greater symmetry.  The idea is to start at small values of $r$, near the RR source, and integrate outward.  

Even without assuming large $k$, the flat spacetime brane solution~\eqref{harmonic} will hold when the distance from the source is small compared to all other scales.  This provides the starting point for the integration.  This solution contains four integration constants, $a, b, c, r_\ast$.  Two are fixed by the equations of motion and a coordinate choice as before, 
\begin{equation}\label{parameters} 
a = c = \frac{2}{\sqrt{k}}\ .
\end{equation}
The remaining two are again fixed by fitting onto the asymptotic geometry.  

The way this works in the numerical integration is that generic choices of $b$ and $r_\ast$ do not lead to the $\text{AdS}_{5} \times {\bf S}^{5}/\mathbb{Z}_{k}$ geometry asymptotically.  To see this, let us linearize the Einstein equations~(\ref{EOM}) around the asymptotic solution~(\ref{BG}), defining
\begin{equation}
\rho = 1 + \omega\ ,\quad f = \frac{R^2}{4}e^{2r/R}(1 +\phi)\ ,\quad g = R^2(1 + \gamma)\ ,\quad
h = R^2(1 + \zeta)\ .
\end{equation}
The $E_g $ and $E_h $ equations reduce to
\begin{equation}
\left(\partial_r^2 + \frac{4}{R} \partial_r \right) \left[ \begin{array}{c} \gamma \\ \zeta 
\end{array} \right]
= \frac{1}{R^2}
 \left[ \begin{array}{cc} 28 & 4 \\ 16 & 16 
\end{array} \right]
 \left[ \begin{array}{c} \gamma \\ \zeta 
\end{array} \right]\ ,
\end{equation}
with solutions going as $e^{4r/R}, e^{2r/R}, e^{-6r/R}, e^{-8r/R}$.  The perturbation~$\omega$ is a gauge choice, and the perturbation $\phi$ is fixed in terms of the others by the first order $E_\rho$ equation.  Thus there are two nonnormalizable modes.   

Generic values of the parameters at small $r$ will lead to nonzero coefficients for these modes at large $r$, but we have two parameters to adjust there, so there is exactly the right amount of freedom to match onto the desired $\text{AdS}_{5} \times {\bf S}^{5}/\mathbb{Z}_{k}$ asymptotic geometry.  We do not need to hunt blindly in the $(b,r_\ast)$ plane because we already know the matching values~(\ref{match}) in the large $k$ limit.  Thus we parameterize 
\begin{eqnarray}
b_k &=& \frac{\sqrt{k}}{2} +  \beta_k \ ,\nonumber\\
\ln (r_{k\ast}/r_0) &=&   \frac{k}{4} + \delta_k\ , \label{betdelt}
\end{eqnarray}
and expect $\beta_k$ and $\delta_k$ to be order unity at large $k$.  Once we find the large-$k$ solution we step down to smaller values.

We fix the coordinate freedom by the choice
\begin{equation}\label{gauge}
g(r) = R^2\rho(r)\ ,
\end{equation}
which conveniently fits both the large-$r$ and small-$\tilde r$ forms~\eqref{BG} and \eqref{harmonic} (thus we can drop the tilde).  To be precise, this gauge choice leaves a freedom $r \to r + v$, so the metric obtained by numerical integration to large-$r$ will differ from the form~(\ref{BG}) by a translation of the coordinate.

\begin{table} \label{table1}
\vspace{.2in}
    \begin{center}
    \begin{tabular}{| l | l | l |}
    $\ \ k$ & $\quad\quad\beta$ & $\quad\quad\delta$ \\
    \hline
    $ 10$  & $ 0.997040$ & $  -0.37062038$ \\
    $ 9$  & $  1.100418$ &$  -0.38625573$ \\
    $ 8$  & $  1.239230$ &$  -0.40796333$ \\
    $ 7$  & $  1.438662$ &$  -0.43836289$ \\
    $ 6$  & $  1.757015$ &$  -0.48153266$ \\   
    $ 5$  & $  2.368251$ &$  -0.54412993$ \\
    $ 4$  & $  4.139490$ &$  -0.63777863$ \\   
    $ 3.9$  & $  4.528660$ &$  -0.64956447$ \\   
    $ 3.7$  & $  5.637135$ &$  -0.67480935$ \\   
    $ 3.5$  & $  7.625691$ &$  -0.70252395$ \\   
    $ 3.3$  & $  12.253897$ &$  -0.73304036$ \\  
    $ 3.1$  & $  35.35767$ &$  -0.76675440$ \\       
    $ 3.01$  & $  347.1317$ &$  -0.78309243$ \\       
    $ 3.001$  & $  3464.86$ &$  -0.78476896$ \\               
    \hline
    \end{tabular}
    \caption{Values of the parameters~(\ref{betdelt}) found numerically.}
    \end{center}
    \end{table}
 In Table 1 we give the value of the parameters for various values of $k$.  By adjusting these parameters
 to one part in $10^6$, the larger growing mode $e^{ 4r/R}$ remains small down to around $r \sim 4R$.  For example, Fig.~1 shows $g_{\chi\chi}/R^2$ out to  $r \sim 2R$ for $k=10$, with good convergence to the asymptotic  $\text{AdS}_{5} \times {\bf S}^{5}/\mathbb{Z}_{k}$ value~1; other metric components show similar convergence.  Thus, the numerical integration carries well into the region where the linearized analysis above is valid, and we can conclude that these are good solutions.
 
  \begin{figure}[h]
\centering
\begin{picture} (0,0)
    	\put(-130,5){$g_{\chi \chi}/R^2$}
	      \put(145, -170){$\frac{r}{R}$}
    \end{picture}

	\includegraphics[scale = 1]{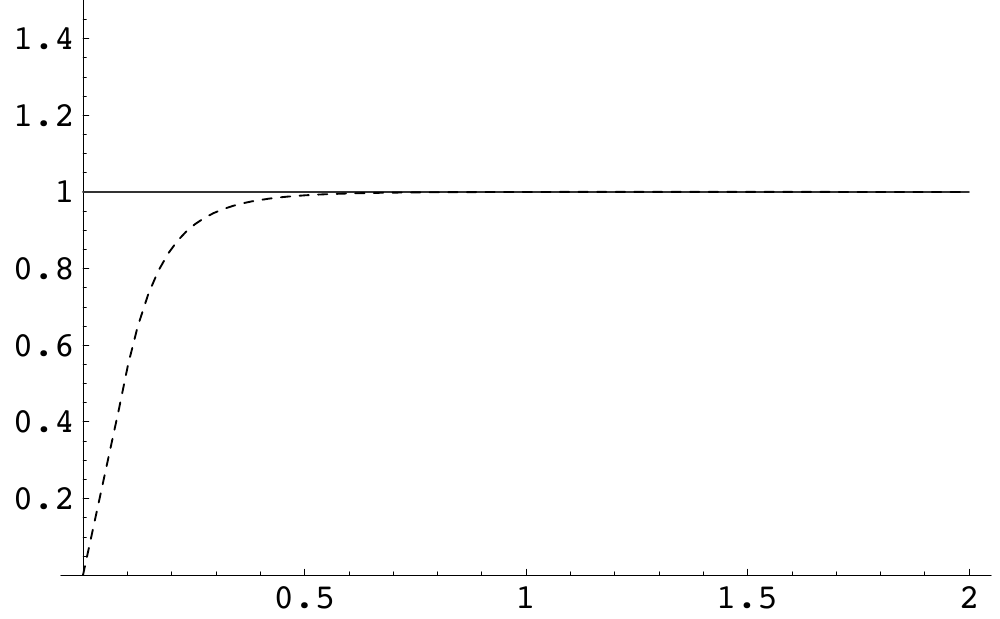}
	
	\caption{$g_{\chi \chi}$ for the $k = 10$ bounce (dashed line) and the asymptotic form (solid line). }
\end{figure}
 
The parameter $k$ does not appear in Einstein's equations but only in the initial conditions~(\ref{harmonic}).  In the full problem it must be an integer or else the metric has a singularity, but at the level of the differential equations $k$ might take any real value.  It is evident from Table~1 that a solution exists for all $k > 3$ but that the solution diverges as we try to take $k$ to 3.  Thus, all nonsupersymmetric cases are unstable, but there is no bounce solution in the supersymmetric case $k = 3$.  This is consistent with the spirit of the positive energy theorems \cite{Gibbons:1983aq}, which show that supersymmetric vacua cannot decay. However, the existing theorems require a nonsingular spacelike surface, so they do not directly apply to this case with D-brane sources.  

\setcounter{equation}{0}

\section{Discussion}

What is the consequence of this instability for the bulk theory and the dual field theory?  The decay rate per unit volume in the bulk will be $e^{-B}$ times a dimensional prefactor.  For the localized bounce the prefactor is dimensionally of order $r_0^{-10}$.  For the smeared bounce there will be additional factors of $r_0/R \sim 1/k$, but we will not need this precision; in any case the localized decay is faster.  The total decay rate per unit gauge theory four-volume is then\footnote{The gauge theory naturally lives on a sphere of radius $R$.}
\begin{equation}
\Gamma \sim \frac{e^{-B}}{r_0^{10} R^4} \int dr \, d^5\Omega\, \sqrt{g_{10}} \sim \frac{R^{2} e^{-B}}{k r_0^{10}} \int_0^\infty dr\, r^3\ ;  \label{div}
\end{equation}
this is in a coordinate such as (\ref{ads5s5}) where the warp factor is linear in $r$ at large $r$.  This rate diverges at the boundary, that is, in the UV.  This is as it must be: in a conformally invariant theory a rate must either be zero or infinite.  Thus the field theory decays immediately.  The bulk is completely consumed by the expanding bubbles, leaving no degrees of freedom at all.\footnote{Note however that a given region of bulk spacetime may survive for almost an AdS time before being overtaken by an expanding bubble from near the boundary.  We thank Tom Banks for raising this point.}

If we cut the field theory off at an energy scale $\Lambda$, this corresponds to cutting off the radial coordinate at $r_{\rm c} \sim R^2 \Lambda$~\cite{SussWit}, and so
\begin{equation}
\Gamma \sim k^{9} e^{-B} \Lambda^4\ .
\end{equation}
Since the field theory lives on a sphere of radius $R$,  the decay rate per unit time is $\Gamma R^3$. We can thus think of the field theory as defined in an effective sense, over spacetime scales long compared to $\Lambda^{-1}$ and short compared to $(\Gamma R^3)^{-1} = \Lambda^{-1}e^{B}/  (k^3\Lambda R)^3$.  Since $B \sim N^2/k^8$, this ratio can be quite large in the 't Hooft limit of large $N$ with $k$ fixed.  Over longer times, the decay takes place and the bulk spacetime disappears for $r < r_{\rm c}$.

Thus, the strongly and weakly coupled theories are both effective theories requiring a UV cut-off  and exhibiting IR instabilities. 
The nature of the instability in the strongly coupled gauge theory can be inferred from the geometry.  The pinching off of the KK circle means that winding strings have vacuum expectation values, because winding number is no longer conserved.  In the gauge theory this corresponds to an expectation value of twisted-trace operators of the form 
\begin{equation}\label{tt}
{\rm Tr}(g'^p {\cal O})\ ,
\end{equation}
where $g'$ is given in (\ref{gprime}) and ${\cal O}$ is any product of the fields.  This is the same operator that is expected to get a vacuum expectation value in the weakly coupled gauge theory~\cite{DFKR}.  Refs.~\cite{Dymarsky:2005uh,DFKR} consider both $SU(3)$ singlet and adjoint ${\cal O}$'s, which appear to correspond to  the localized and smeared bounces, respectively, and find instabilities in both kinds of operator.
Refs.~\cite{Armoni:2003va,DFKR} suggests that even after the condensation, a massless ${\cal N} = 4$ sector remains.  This is evidently mirrored by our observation that the decay will in general lead to the production of explicit D3-branes.\footnote{In our notation, the parent gauge group is $SU(N)$ and $N/k$ D3-branes are produced.  In the gauge theory papers~\cite{Adams:2001jb,Dymarsky:2005uh,Dymarsky:2005nc,DFKR} the parent gauge theory is $SU(Nk)$ and $N$ D3-branes would be produced.}

Although the symmetry breaking patterns are similar at weak and strong coupling, the mechanisms that drive the breaking are different.\footnote{As a rough analogy consider superconductivity, where the same symmetry breaking is driven by phonon exchange in ordinary superconductors, and electronic effects in high-$T_c$ superconductors.  The Casimir instability that sets in at $k^{10} \sim N^2$, as discussed in the introduction, is another mechanism that drives the same breaking in the orbifold.}
  At weak coupling it is due to running of the double twisted-trace couplings (coefficients of the product of operators like (\ref{tt})) already at planar order, so that they become strong at low energy.  As has been remarked beginning in Ref.~\cite{Adams:2001jb}, there is no such running at strong coupling: the twisted-trace two-point functions are conformal in the classical bulk theory.  This is not surprising: the beta functions for some of the double twisted-trace couplings have no real zeros at small $\lambda$~\cite{Dymarsky:2005uh,Dymarsky:2005nc}, but such zeros might well appear as $\lambda$ is increased, and the evidence from the bulk side is that this is the case.  Thus, the symmetry breakdown at strong coupling appears to be driven directly by large $\lambda$.  This is consistent with the fact, evident in Eq.~(\ref{div}), that the instability is present at all scales.  (Ref.~\cite{Coleman:1974bu} presents an interesting prototype of a strongly coupled conformal theory that is unstable at all scales).  It is conceivable that at intermediate values of $\lambda$ the twisted beta functions develop a zero before the large-$\lambda$ KK instability appears, so there would be a phase without symmetry breaking.

The 't Hooft parameter $\lambda$ runs only at non-planar order $1/N^2$.  At weak coupling this is slow compared to the planar running of the double twisted couplings and can be neglected.  At strong coupling one might think at first that this perturbative correction dominates the exponentially suppressed nonperturbative decay, but it does not because the running is logarithmic in the energies.  Further, there is an additional suppression that makes the running unimportant.  The supergravity mode frequencies in Planck units are independent of the dilaton, and so do not generate a dilaton tadpole.  The supergravity Casimir energy has the same dilaton dependence as the five-form flux energy, with an additional factor of $k^{10}/N^2$, and so produces only a small shift of the geometry.  The leading dilaton dependence would come through string corrections to the low energy action, and so would be suppressed by additional powers of $\alpha'/R^2 = \lambda^{-1/2}$.  The $\beta$-function is thus of order $N^{-2} \lambda^{-1/2}$, and so the running of the dilaton over the maximum range of scales $O(e^B)$ allowed by the decay is $B N^{-2} \lambda^{-1/2}\sim \lambda^{-1/2}$ (times some power of $k$), becoming negligible at large $\lambda$.

We have shown that $\text{AdS}_{5} \times {\bf S}^{5}/\mathbb{Z}_{k}$ (for $k>3$)  has a nonperturbative instability analogous to the decay of the Kaluza-Klein vacuum.
In light of this, it is natural to ask if the total energy is bounded from below for all supergravity solutions with $\text{AdS}_{5} \times {\bf S}^{5}/\mathbb{Z}_{k}$ boundary conditions.  There is clearly no analog of the usual positive energy theorem since the instantons describe tunneling from $\text{AdS}_{5} \times {\bf S}^{5}/\mathbb{Z}_{k}$ to nontrivial solutions with exactly the same energy. So the zero energy solution is not unique. For standard Kaluza-Klein boundary conditions, it has been shown that solutions exist with  arbitrarily negative energy \cite{Brill:1989di,Brill:1991qe}. It is very likely that the same is true here. Since a solution with an accelerating large bubble has zero energy, one expects that by decreasing the kinetic energy, one could find solutions with arbitrarily negative energy.

An interesting extension of our work would be to the nonconformal conifold theories~\cite{Klebanov:1999rd,Klebanov:2000nc,Klebanov:2000hb}.  Here the effective value of $N$ appearing in the AdS radius grows as $\ln r$, so the bounce amplitude $e^{-B}$ decreases faster than a power of $r$, and the total rate~(\ref{div}) converges.  Correspondingly, when bubbles do form at small $r$, their growth is likely to be limited by the expansion of the KK circle as in Ref.~\cite{Aharony:2002cx}.  In this case, then, the decay is only an IR effect as it is at weak coupling, and there should be a static geometry with a bubble of finite size.  It would be interesting to verify these geometric expectations.

\vskip 1cm
\centerline{\bf Acknowledgements}
We thank T. Banks, E. Lopez, R. Roiban and E. Silverstein  for discussions.
This work was supported in part by NSF grants PHY05-55669, PHY05-51164 and PHY04-56556.

\end{document}